\documentclass[prb,preprint,groupedaddress,amsmath,amssymb]{revtex4-1}
\usepackage{amsmath}
\usepackage{physics}
\usepackage{amsfonts}
\usepackage{bm}
\usepackage{graphicx}
\usepackage{hyperref}
\usepackage{pifont}
\usepackage{color}
\usepackage{tabularx}

\begin{document}

\title{Valley-Filling Instability and Critical Magnetic Field for Interaction-Enhanced Zeeman Response in Doped WSe$_2$ Monolayers}

\author{Fengyuan Xuan}
\affiliation{Centre for Advanced 2D Materials, National University of Singapore, 6 Science Drive 2, Singapore 117546}
\author{Su Ying Quek}
 \email{phyqsy@nus.edu.sg}
\affiliation{Department of Physics, National University of Singapore, 2 Science Drive 3, Singapore 117551}
\affiliation{Centre for Advanced 2D Materials, National University of Singapore, 6 Science Drive 2, Singapore 117546}
\affiliation{NUS Graduate School Integrative Sciences and Engineering Programme, National University of Singapore}
\affiliation{Department of Materials Science and Engineering, National University of Singapore}

\date{\today}

\begin{abstract}
Carrier-doped transition metal dichalcogenide (TMD) monolayers are of great interest in valleytronics due to the large Zeeman response (g-factors) in these spin-valley-locked materials, arising from many-body interactions. 
We develop an \textit{ab initio} approach based on many-body perturbation theory to compute the interaction-enhanced g-factors in carrier-doped materials. We show that the g-factors of doped WSe$_2$ monolayers are enhanced by screened exchange interactions resulting from magnetic-field-induced changes in band occupancies. Our interaction-enhanced g-factors $g^*$ agree well with experiment. 
Unlike traditional valleytronic materials such as silicon, the enhancement in g-factor vanishes beyond a critical magnetic field $B_c$ achievable in standard laboratories. 
We identify ranges of $g^*$ for which this change in g-factor at $B_c$ leads to a valley-filling instability and Landau level alignment, 
which is important for the study of quantum phase transitions in doped TMDs. 
We further demonstrate how to tune the g-factors and optimize the valley-polarization for the valley Hall effect.
\end{abstract}

\maketitle
Valleytronics, the control and manipulation of the valley degree of freedom (valley pseudospin),
is being actively considered as the next new paradigm for information processing.
The field of valleytronics dates back to investigations on traditional semiconductors such as silicon\cite{Ohkawa,Sham},
but the ability to exploit valley polarizations in these materials has been limited \cite{NatRevMat}.
A major impetus for the renaissance of valleytronics is the recent discovery that H-phase transition metal dichalcogenide (TMD) semiconductor monolayers (MLs)
are excellent candidates for valleytronics applications \cite{NatRevMat,RMP2018}.
The spin-valley locking effect in these MLs \cite{Xiao2012} leads to long lifetimes for spin- and valley-polarization, 
while individual valleys can be probed and controlled using circularly-polarized light, 
paving the way to use the valley pseudospin for information processing.
The valley Zeeman response in TMD MLs \cite{Xuan2020,Rohlfing2020,PRB2020,PRB2020,NC2020,Marie2020,2018Ensslin,LL2019,Mak,2020PRL,MakPRL,jump,NanoLett,NatMater,LL2017,2018PRBR}
is also significantly larger than in traditional semiconductors \cite{Roth,Yafet,Janak,Fang,Lak,ExpPRB,Shayegan}.

When an external magnetic field is applied normal to the TMD ML, the energies of the valleys shift in equal magnitude and opposite directions.
This Zeeman effect is quantified by the orbital and spin magnetic moments, which contribute to the Land\'{e} g-factors.
In TMD MLs, the intrinsic Land\'{e} g-factors are about $6$ times larger \cite{Xuan2020,Rohlfing2020,PRB2020,Marie2020} than that in silicon,
where only the spin magnetic moment dominates \cite{Roth,Yafet}.
The larger g-factors in TMD MLs allow for greater control in tuning the energetics of the valley pseudospins, 
and results in a larger valley-polarized current, which is important for observations of the valley Hall effect \cite{NatRevMat}.
Besides the Zeeman effect, an external magnetic field also results in a quantization of states to form Landau levels (LLs).

Much of the current research on valleytronics seeks to understand how to manipulate the valley pseudospins in TMDs \cite{NatRevMat,RMP2018}.
It has been found that carrier doping can dramatically enhance the g-factors in TMD monolayers \cite{2020PRL,MakPRL,jump,NanoLett,NatMater,LL2017,2018PRBR},
opening up the possibility to tune the valley pseudospin by gating in a magnetic field.
This enhancement in g-factors has been attributed to many-body interactions \cite{2020PRL,MakPRL,jump,NanoLett,NatMater,LL2017,2018PRBR,Janak,ExpPRB}, but a quantitative understanding is lacking.
To interpret the g-factor enhancement in doped TMDs, experimentalists have typically relied on existing theoretical literature dating back to the 1960s-1970s\cite{Janak,Ando}. However, there are two shortcomings of these theoretical approaches. Firstly, they are not \textit{ab initio} methods and cannot provide quantitative predictions. Secondly, these studies all focused on silicon or III-V semiconductors, which are very different in nature from the TMD monolayers. It is important to question if the experimental observations on g-factors in doped TMDs serve only to validate in a new material what was already known for silicon,
or if it is possible to observe novel phenomena not known before in conventional valleytronics materials.

In this work, we develop an \textit{ab initio} approach based on many body perturbation theory to compute the interaction-enhanced Land\'{e} g-factors in carrier-doped systems. We predict that the larger intrinsic g-factors in TMD MLs enable the observation of a critical magnetic field $B_c$
above which the interaction-induced enhancement in the g-factors vanishes in doped TMDs.
We identify ranges of the enhanced g-factor $g^*_{\text{enh}}$ for which the discontinuous change in g-factor at $B_c$ results in a LL alignment and valley-filling instability for $B \gtrsim B_c$.
Such a phenomenon has not been observed or predicted for silicon and other conventional valleytronic materials. 
Our computed interaction-enhanced g-factors for hole-doped ML WSe$_2$ agree well with experiment \cite{2020PRL,jump,NatMater} and can be tuned by dielectric screening.
The predicted valley-filling instability for $B \gtrsim B_c$ provides theoretical insights into recent experimental observations \cite{jump} of a pronounced Landau level-filling instability at a critical magnetic field which closely matches our predicted values. 
The associated alignment of LLs is of interest\cite{jump} to investigate quantum phase transitions in these doped TMDs \cite{Braz,Donk,Miserev,Roch,Nature1999,Science2000}.
The recent observation of fractional quantum Hall states associated with non-abelian anyons in ML WSe$_2$ \cite{FQHWSe2} highlights the potential of creating pseudo-spinors from aligned LLs for topological quantum computing applications \cite{QCRMP}.

\textbf{RESULTS}

\textbf{Interaction-enhanced g-factors}

For a many-electron system described by a static mean-field Hamiltonian $H \ket{n\mathbf{k}} = E_{n\mathbf{k}}\ket{n\mathbf{k}}$,
it has been shown that an out-of-plane magnetic field $B$ results in the following expression for the LLs at the $K$ valley \cite{Xuan2020}:
\begin{eqnarray}
\epsilon_{N,K} = E_{n\mathbf{K}}+(N+\frac{1}{2})\frac{2}{m^*}\mu_B B - g^{I}_{n\mathbf{K}} \mu_B B,
\label{eLL}
\end{eqnarray}
where $n$ is the corresponding band index, $E_{n\mathbf{K}}$ is the mean-field single-particle energy at $K$, 
$m^*$ is the valley effective mass, $\mu_B$ is the Bohr magneton and $N = 0, 1, 2, ...$ is the LL index.
The total intrinsic single band g-factor consists of the orbital and spin contribution,  $g^{I}_{n\mathbf{K}}=g^{\text{orb}}_{n\mathbf{K}}-g_s s_{z,n\mathbf{K}}$, where $g_s$ is taken to be $2.0$ and $s_{z,n\mathbf{K}}$ is the spin quantum number.
The orbital g-factor is defined using the orbital magnetic moment $g^{\text{orb}}_{n\mathbf{K}} \mu_B = m^z_{n\mathbf{K}}$ \cite{Xuan2020} (see Methods).

However, the above static mean-field description does not account for the energy-dependent electron self-energy $\Sigma(E)$ 
that is necessary for a many-body description of the quasiparticle (QP) energies. 
$\Sigma(E)$ represents the change in the energy of the bare particle due to the interaction of the particle with itself \textit{via} the interacting many-body system. 
The change in self-energy as the QP energy shifts with $B$ leads to an effective renormalized g-factor, 
$g^*_{n\mathbf{K}}$, defined as $(E_{n\mathbf{k}}^{\text{QP},1}-E_{n\mathbf{k}}^{\text{QP}})=g^*_{n\mathbf{K}} \mu_B B$, 
where 
\begin{eqnarray}
\begin{split}
E_{n\mathbf{k}}^{\text{QP}} = E_{n\mathbf{k}} & + \Sigma(E_{n\mathbf{k}}^{\text{QP}}) \\
E_{n\mathbf{k}}^{\text{QP},1} = E_{n\mathbf{k}} & + \Sigma(E_{n\mathbf{k}}^{\text{QP},1}) + g^I_{n\mathbf{K}} \mu_B B,
\label{EQP1}
\end{split}
\end{eqnarray}
Thus we have,
\begin{eqnarray}
\frac{g^{I}_{n\mathbf{K}}}{g^*_{n\mathbf{K}}} = 1 - \frac{d\Sigma(E)}{dE}.
\label{effg}
\end{eqnarray}
Such a renormalization effect is missing in previous first principles calculations of g-factors in TMDs \cite{Xuan2020,Rohlfing2020,PRB2020}.

\begin{table}[h!]
\caption{
\textbf{$GW$ single band and exciton g-factors of undoped ML WSe$_2$ at the $K$ valley.} 
$X0$ and $D0$ refer to the lowest energy spin-allowed and spin-forbidden optical transitions \cite{Xuan2020,Rohlfing2020,PRB2020}.
For the single-band g-factors, $c$ and $v$ refer to the frontier conduction and valence bands, respectively,
while $\uparrow$ and $\downarrow$ refer to spin up and spin down bands at $K$.
}
\begin{ruledtabular}
\begin{tabular}{l*{2}{c}}
           & Renormalized ($g^*$) & Intrinsic ($g^{I}$)\\
\hline
$g_{c\uparrow}$          &  -4.36 & -5.50    \\
$g_{c\downarrow}$        &  -2.05 & -2.59    \\
$g_{v\uparrow}$          &  -6.63 & -8.31   \\
$g_{v\downarrow}$        &  -4.48 & -5.62   \\
\hline
$g_{X0}$                 &  -4.54 & \\
$g_{D0}$                 &  -9.16 & \\
\end{tabular}
\label{tabg}
\end{ruledtabular}
\end{table}

The electron self-energy in this work is computed within the $GW$ approximation \cite{Hedin} (see Methods),
which uses the first order term in the perturbative expansion of $\Sigma$ in terms of the screened Couloumb interaction $W$.
Table~\ref{tabg} shows the renormalized and intrinsic g-factors computed for undoped monolayer WSe$_2$. 
We see that the magnitudes of the renormalized g-factors are reduced by $\sim20$\% compared to the intrinsic $GW$ g-factors, 
because $\frac{\partial\Sigma(E)}{\partial E}$ is in general negative \cite{Louie}. 
The exciton g-factors deduced using the renormalized g-factors are in good agreement with experiment  \cite{2d2015,NC2019,NatPhys2015,2d2019,Liu2019}. 

\begin{figure*}[h!]
\centering
\includegraphics[width=17.0cm,clip=true]{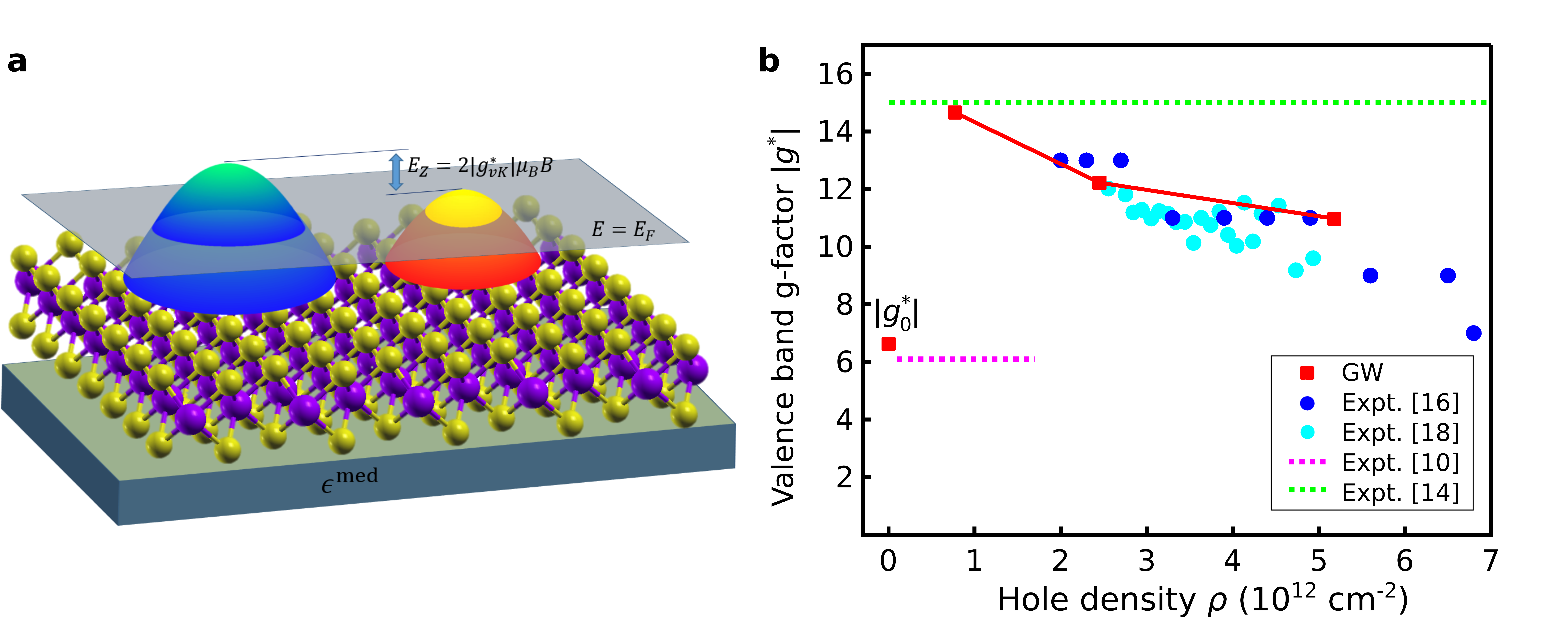}
\caption{
\textbf{Valence band g-factor in hole-doped ML WSe$_2$.} 
\textbf{a} Schematic figure of the energy dispersion of hole-doped WSe$_2$ ML in the presence of an out-of-plane magnetic field, in the mixed polarized regime where the g-factor is enhanced. 
\textbf{b} Valence band g-factor $|g^*|$ in hole-doped ML WSe$_2$ as a function of hole density.
Red squares: Calculated results; Dark blue and light blue circles: Experimental data from Ref. \onlinecite{jump}, \onlinecite{NatMater}, respectively;
Purple and green dotted lines: Experimental data from Ref. \onlinecite{Marie2020}, \onlinecite{2020PRL}, respectively (hole densities are given in a range only).
}
\label{gstarfig}
\end{figure*}

In contrast to the undoped system, doped systems have partially occupied bands. 
Thus, if the band occupancies are also changing in response to the magnetic field, there is an additional term in $\frac{d\Sigma(E)}{dE}$:
\begin{eqnarray}
\frac{d\Sigma(E)}{dE} = \frac{\partial\Sigma(E)}{\partial E}+\frac{\partial\Sigma(E)}{\partial f}\frac{\partial f}{\partial E}
\label{dSdE}
\end{eqnarray}
where $f$ is the Fermi-Dirac distribution function. 
The second term in Eq~(\ref{dSdE}) can be simplified to give (see Methods):
\begin{eqnarray} 
\begin{split}
\frac{\partial\Sigma(E)}{\partial f}\frac{\partial f}{\partial E} \approx \frac{\lvert m^* \rvert}{2\pi} \bar{W}_{nk_F}(E=E_F),
\label{enh_main}
\end{split}
\end{eqnarray}
where $n$ is the band index of the frontier doped band, and $E_F$ and $k_F$ are respectively the Fermi energy and Fermi wave vector. 
This term comes from the screened exchange contribution to the self-energy.
$\bar{W}_{nk_F}$ (defined in Methods) is an effective quasi-2D screened Coulomb potential which can be evaluated completely from first principles.
Since $\bar{W}_{nk_F}$ is positive, the second term of Eq.~(\ref{dSdE}) leads to an enhancement effect for the g-factor. 
For an ideal 2D fermion gas, this second term reduces to the term $d\Sigma(E)/dE$ derived in Ref. \onlinecite{Janak} (see also Ref. \onlinecite{Quinn}) 
where the first term in Eq.~(\ref{dSdE}) is ignored. 
We note that our numerical results for $\bar{W}_{vk_F}$ as a function of hole density differ from those computed for an ideal 2D fermion gas\cite{stern,Janak}, although the corresponding numerical results for the bare Coulomb potential match well with the ideal 2D case  (see Supplementary Figure 1). This observation implies that an \textit{ab initio} non-local description of the dielectric function of the quasi-2D system is important for a quantitative prediction of the renormalized g-factors.   

The g-factors in this work are all computed for the valence band at $K$, and henceforth, the subscript $v\mathbf{K}$ is omitted. The magnitudes of our computed valence band g-factors $|g^*|$ are plotted as red squares in Fig.~\ref{gstarfig}b for ML WSe$_2$ with different hole densities. 
Our predicted g-factors agree well with those deduced from multiple experiments \cite{jump,2020PRL,Marie2020,NatMater} on hole-doped ML WSe$_2$. 
The renormalized g-factor for the undoped system is labeled $g^*_0$.
Due to interaction-induced enhancement, the g-factor increases significantly once hole carriers are introduced. 
This enhancement reduces as the hole density is increased as expected from the density dependence of the many-body Coulomb interactions (see Supplementary Figure 1).

\textbf{Critical magnetic field}

\begin{figure*}[h!]
\centering
\includegraphics[width=17.0cm,clip=true]{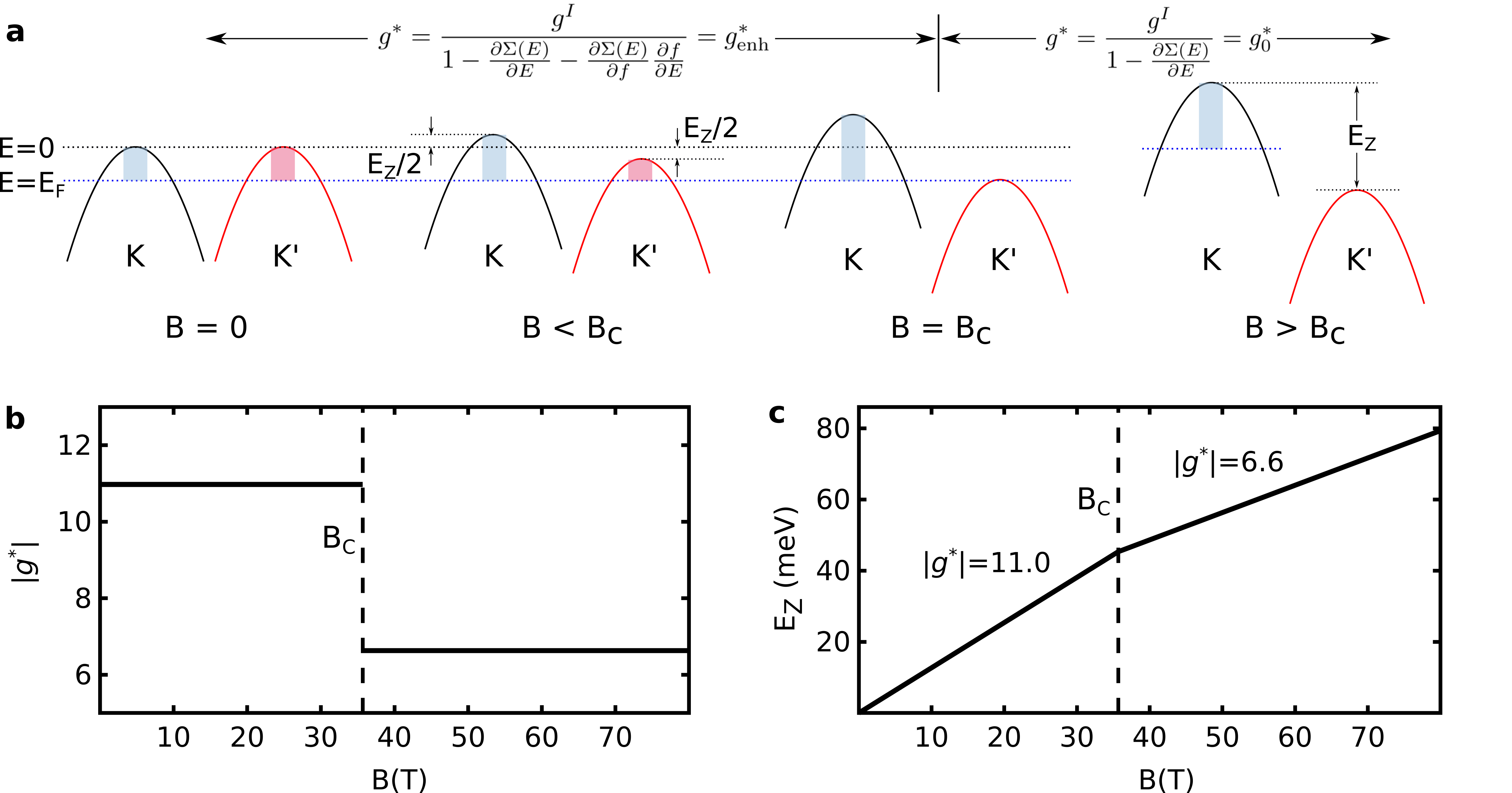}
\caption{
\textbf{Magnetic field dependence of g-factor.}
\textbf{a} Schematic figure illustrating the Zeeman effect as $B$ is increased with fixed hole density. 
The pink and blue shaded regions illustrate the magnitude of hole density due to the constant density of states in the quadratic band in two dimensions. 
As $B$ increases for $B \leq B_c$, the absolute Fermi level position is unchanged, but the band occupancies change. 
For $B>B_c$, the Fermi level shifts with the band at $K$ and the band occupancies remain unchanged.
\textbf{b-c} Plots of \textbf{b} $|g^*|$ and \textbf{c} valley Zeeman split $E_Z$ (labeled in \textbf{a})  
as a function of magnetic field $B$ for monolayer WSe$_2$ with hole density $5.2\times 10^{12}\text{cm}^{-2}$.
}
\label{schfig}
\end{figure*}

A subtle but important point is that Eq.~(\ref{enh_main}) applies only when the band occupancies are changing with $B$, and the Fermi level $E_F$ is fixed.
In electrostatic gating experiments, the carrier concentration is fixed rather than the absolute Fermi level.
However, the Zeeman shifts in $K$ and $K'$ are equal in magnitude but opposite in sign (Fig.~\ref{schfig}a, $B < B_c$), and the density of states for the quadratic bands in 2D is independent of energy.
So for $B$ small enough that both valleys have carriers (mixed polarized regime; Fig.~\ref{gstarfig}a), 
$E_F$ is fixed while the band occupancies change and both terms in Eq.~(\ref{dSdE}) will apply, leading to the interaction-enhanced g-factors, which we label as $g^*_{\text{enh}}$.
However, above a critical magnetic field $B_c \sim |E_F|/(|g^*_{\text{enh}}|\mu_B)$, only one valley has carriers (Fig.~\ref{schfig}a, $B>B_c$) 
(see Supplementary Note for a more precise expression for $B_c$). 
As $B$ increases beyond $B_c$, a constant hole density is maintained when $E_F$ shifts with the bands without changing the band occupancies. 
Thus, for $B>B_c$, only the first term in Eq.~(\ref{dSdE}) applies, similar to the undoped case, leading to an abrupt drop in $g^*$ at $B=B_c$ (Fig.~\ref{schfig}b), 
with a corresponding piecewise-linear Zeeman split $E_Z$ (Fig.~\ref{schfig}c). 
For a hole density of $5.2\times 10^{12}\text{cm}^{-2}$, $B_c \approx 36$ T, and we have $g^*\approx-11.0$ for $B<B_c$ and  $g^*\approx-6.6$ for $B>B_c$.

This abrupt drop in $g^*$ at a critical magnetic field has never been reported or predicted before in traditional valleytronic materials such as silicon. 
Indeed, $B_c$ is inversely related to $|g^*_{\text{enh}}|$, 
and it is the large intrinsic g-factors and hence large renormalized g-factors for TMDs that allow for $B_c$ to be small enough to be reached in standard laboratories. 
For the same hole density of  $5.2\times 10^{12}\text{cm}^{-2}$, we predict $B_c$ in silicon to be $\sim 400$ T. 
The larger intrinsic g-factors for TMD MLs arise from the large orbital g-factors, which consist of a valley term, an orbital term, 
and a cross term that involves coupling between the phase-winding of the Bloch states and the parent atomic orbitals. \cite{Xuan2020} 

\begin{figure*}[h!]
\centering
\includegraphics[width=17.0cm,clip=true]{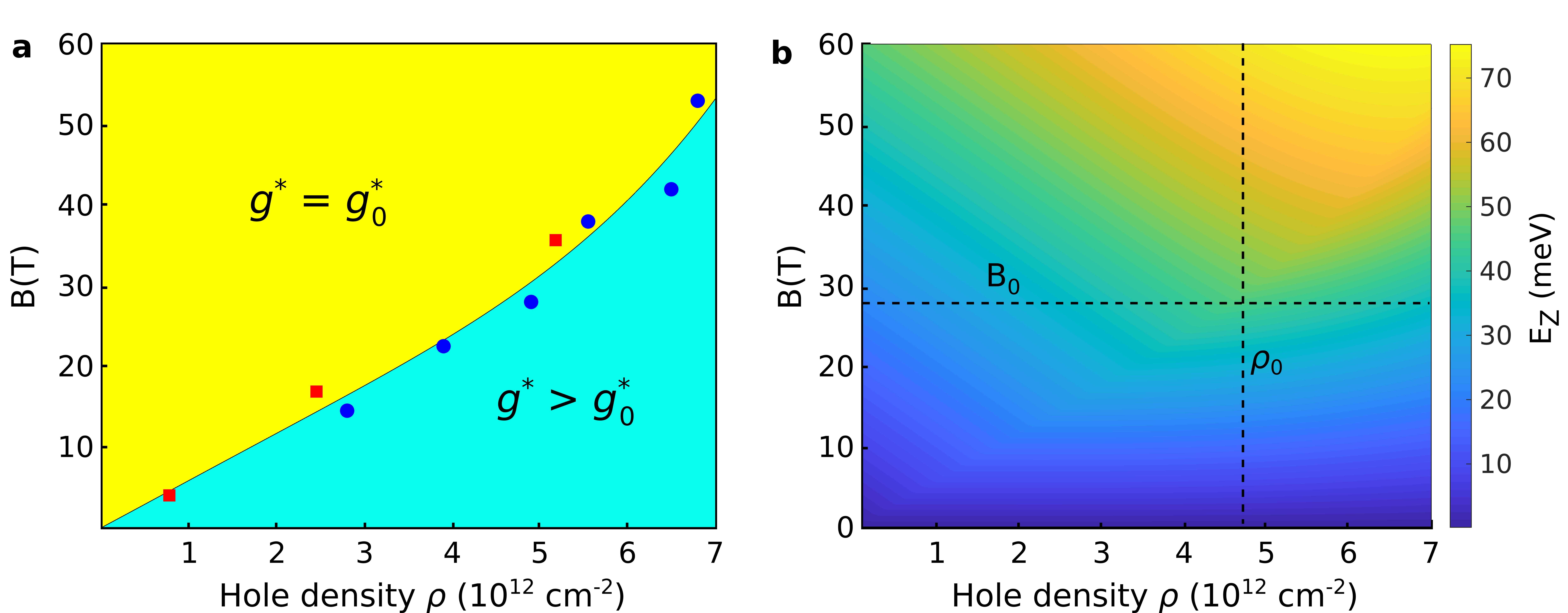}
\caption{
\textbf{Critical magnetic field and Zeeman split.}
(a) Critical magnetic field as a function of hole density. Blue circles: Experimentally derived from the onset of the fully-polarized regime (Ref. \onlinecite{jump}). Red squares: Predicted in this work.
In the yellow-shaded area, the renormalized valence band g-factor is the value of $g^*$ in the undoped TMD ML, $g^*_0$. 
In the cyan-shaded area, the valence band g-factor is enhanced by interactions.
(b) 2D plot for $E_Z$ as a function of hole density and external magnetic field.
}
\label{2dfig}
\end{figure*}

Since $B_c$ is the value of $B$ characterizing the onset of the fully polarized regime, 
$B_c$ can be deduced using optical measurements of the exciton and polaron energies for $K$ and $K'$\cite{jump}. 
In Fig.~\ref{2dfig}a, we plot these values of $B_c$ (blue circles) and compare them with our predicted values (red squares). 
The predicted dependence of $B_c$ on the hole density agrees well with experiment. 

How can one maximize the concentration of valley-(and spin-)polarized carriers in the TMD ML? As the hole concentration increases, 
$B_c$ increases, giving a larger range of $B$ for which $g^*$ is enhanced by interactions (Fig.~\ref{2dfig}a). 
However, the magnitude of $g^*_{\text{enh}}$ decreases as the hole concentration increases (Fig.~\ref{gstarfig}). 
These competing effects imply that for any given $B$ field $B_0$, there is an optimal hole concentration $\rho_0$ which maximizes the Zeeman split $E_Z$ (Fig.~\ref{2dfig}b). 
This optimal hole concentration $\rho_0$ corresponds to the hole concentration for which $B_c = B_0$ (Fig.~\ref{2dfig}), 
and yields a maximum concentration of valley-polarized carriers. 
These predictions are useful for realizations of the valley Hall effect and other applications where a high concentration of valley-polarized carriers is desired.

\textbf{LL Alignment and Valley Filling Instability}
\begin{figure*}[h!]
\centering
\includegraphics[width=17.0cm,clip=true]{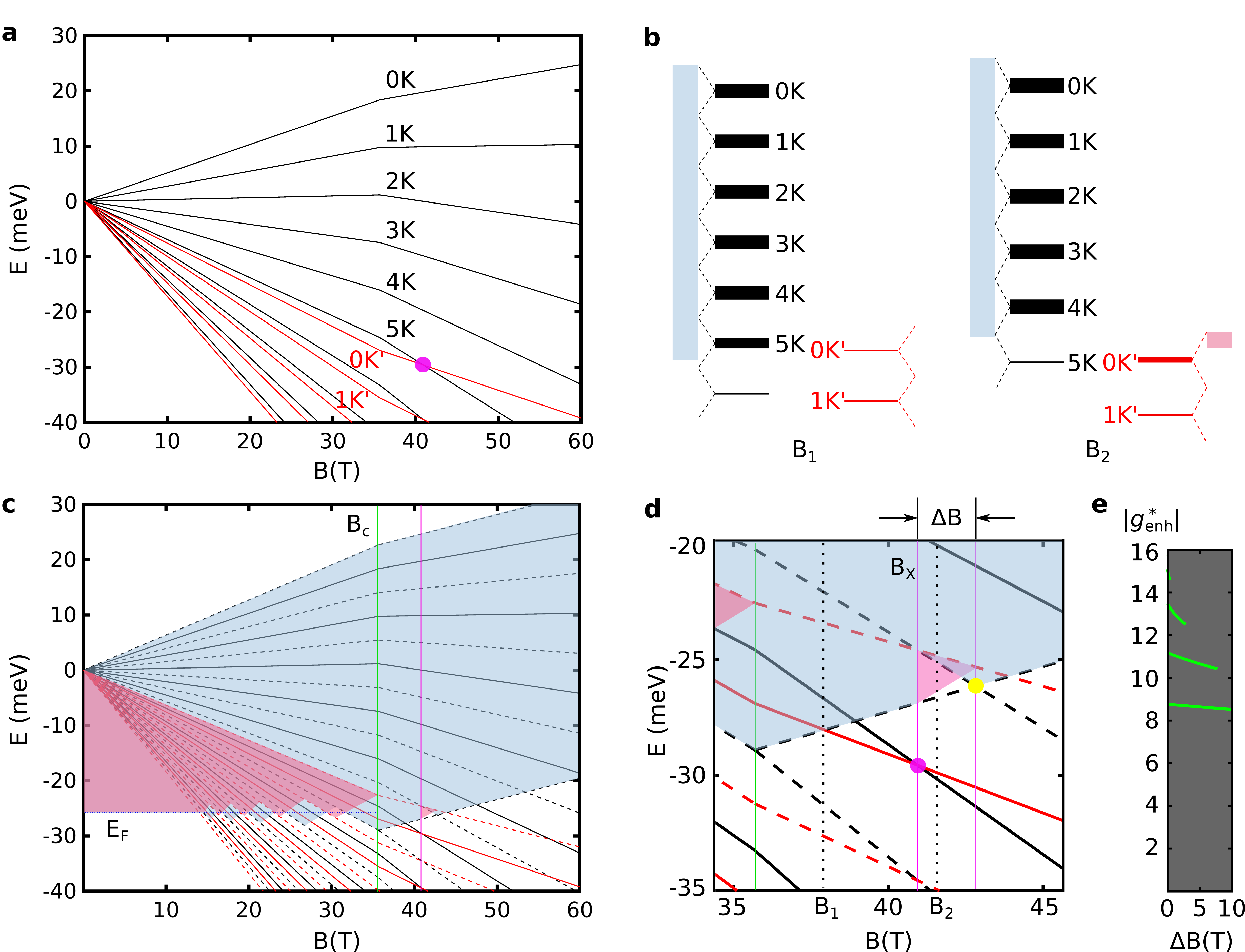}
\caption{
\textbf{LL fan diagram and valley-filling instability.}
\textbf{a} LL fan diagram for ML WSe$_2$ valence band with a hole density of $5.2\times 10^{12}\text{cm}^{-2}$. The LLs are labeled by the LL index $N$ and valley $K$ or $K'$.
The slopes of the plots decrease in magnitude when $B$ increases beyond $B_c$, leading to a crossing between $0K'$ and $5K$ (purple circle).  
\textbf{b} Schematic figure for LLs at $B_1$ and $B_2$ as marked in (d). The blue and pink shading represent the hole populations in $K$ and $K'$, respectively. 
\textbf{c} Hole occupancies of LLs for a constant hole density of $5.2\times 10^{12}\text{cm}^{-2}$. 
The blue/pink shading refer to hole occupations in the nearest LLs, as demarcated by the black/red dashed lines. 
Each LL is fully occupied before the next LL lower in energy, resulting in the zigzag-shaped fine-structure at $E\sim E_F$. 
\textbf{d} Zoom-in figure of (c) showing the valley-filling instability as indicated by the isolated pink triangle. 
The hole population in $5K$ is transferred to $0K'$ for $B>B_X$, until $B$ is large enough (yellow circle) that LLs $0K$ to $4K$ carry all the holes. 
\textbf{e} Range of interaction-enhanced $|g^*_{\text{enh}}|$ for which a valley-filling instability exists, with corresponding $\Delta B$.
}
\label{fullfig}
\end{figure*}

The abrupt change in $g^*$ at $B=B_c$ also has other interesting implications. 
As $B$ increases beyond $B_c$, the decrease in $|g^*|$  results in a decrease in magnitude of the slopes of the LL fan diagrams (Fig.~\ref{fullfig}a), 
leading to a crossing between the energies of $0K'$ and $NK$ (purple circle, $N=5$ in Fig.~\ref{fullfig}a). 
If $NK$ has carriers, this LL alignment results in a valley-filling instability, 
where the hole population is transferred back and forth between the two valleys for small changes in $B$. 

LLs are filled with holes from higher energy to lower energy.
In Fig.~\ref{fullfig}c, the blue and pink shading indicate the filling of the LLs with holes. 
If $NK$ is fully occupied, the plot is shaded from $\epsilon_{NK} + \frac{\hbar \omega_c}{2}$ down to $\epsilon_{NK} - \frac{\hbar \omega_c}{2}$. 
At $B=B_c$, an integer number of LLs ($5$ in Fig.~\ref{fullfig}c) are completely filled with holes while $0K'$ is completely empty. 
As $B$ decreases slightly below $B_c$, the LL degeneracy decreases, and the LL with the next lower energy ($0K'$ in Fig.~\ref{fullfig}c) is required to contain the holes, 
leading to a symmetric zigzag fine structure about $E_F$. 
As $B$ increases slightly above $B_c$, the LL degeneracy increases. 
As long as $\epsilon_{5K}>\epsilon_{0K'}$, only the $K$ valley is filled with holes ($B=B_1$ in Fig.~\ref{fullfig}b), and the blue shaded area represents the constant hole density. 
But when $B$ is slightly larger than $B_X$ (Fig.~\ref{fullfig}d) where the energies of $0K'$ and $5K$ cross, holes will start to fill the $K'$ valley again ($B=B_2$ in Fig.~\ref{fullfig}b), 
until the $B$ field is large enough that the LLs $0K$ to $4K$ can contain all the holes (yellow circle, Fig.~\ref{fullfig}d) and the system becomes fully polarized again. 
This represents a valley-filling instability, where $K'$ is depleted of holes from $B=B_c$ to $B=B_X$, and filled again up to $B_X + \Delta B$. 
In practice, when holes begin to fill $0K'$, the mixed polarized regime is reached and $g^*$ becomes enhanced, 
leading to a change in the slope of the fan diagram that is expected to result in a LL alignment not just for $B=B_X$ but also for $B$ up to $B_X + \Delta B$.

Our predictions provide important theoretical insights into a recent experiment on doped ML WSe$_2$\cite{jump}, where optical absorption plots showed a pronounced signature of the peak positions changing from one inter-LL transition to another over a small range of $B$ close to the onset of the fully-polarized regime in the experiment. 
This is consistent with the highest occupied LL in the $K'$ valley being emptied and partially filled with holes at $B \sim B_c$ in our predictions. 
The authors of Ref. \onlinecite{jump} attributed this observation to the oscillatory g-factors predicted for traditional semiconductors such as silicon \cite{Ando}. 
However, in this theory, the changes in the g-factors are directly related to the position of the Fermi level relative to the LLs, 
and the g-factors therefore have an ``oscillatory"  \cite{Ando} dependence on $B$ rather than a pronounced change at one particular value of $B$ as seen in the experiment. 
Furthermore, such a pronounced instability was not observed in experiments \cite{Fang,Lak,ExpPRB,Shayegan} on the g-factors in doped silicon and other traditional valleytronic materials for which these oscillatory g-factors were predicted. 
Thus, this pronounced instability observed in doped ML WSe$_2$\cite{jump} is in fact a manifestation of the valley-filling instabilities that are predicted here to emerge specifically for doped TMDs. 
Our conclusion is further supported by the fact that the measured values of $B_c$ and $B_X$ are respectively $32$ and $38$ T for $|g^*_{\text{enh}}|\sim 11$\cite{jump}, 
close to our predicted values of $31$ and $37$ T for the same $|g^*_{\text{enh}}|$ (see also Supplementary Table 1). 

The alignment of LLs is of interest to investigate quantum phase transitions in these doped TMDs \cite{jump,Braz,Donk,Miserev,Roch,Nature1999,Science2000}. 
Given that the LLs are expected to align for $B$ between $B_X$ and $B_X + \Delta B$, 
it is interesting to predict how large $\Delta B$ can be and how sensitive $\Delta B$ is to fluctuations in $g^*_{\text{enh}}$. 
Not all values of $g^*_{\text{enh}}$ will result in the instability (see Supplementary Figure 2 and Supplementary Note). In particular, if the energies of $0K'$ and $NK$ cross at $B_X$, the valley-filling instability only occurs if $NK$ is occupied with holes.
Fig.~\ref{fullfig}e plots the ranges of $g^*_{\text{enh}}$ for which a valley-filling instability will occur, as well as the corresponding $\Delta B$ values. 
We see that an optimal range of $|g^*_{\text{enh}}|$ for LL alignment is $10.4<|g^*_{\text{enh}}|<11.2$. 
Here, $\Delta B$ is quite large ($\sim 8$T for $|g^*_{\text{enh}}| \sim 10.4$) and is also fairly robust to changes in $g^*_{\text{enh}}$. 
The corresponding values of $B_c$ and $B_X$ fall within $30$ to $40$ T (Supplementary Table 1), well within the reach of experiments. 

\begin{figure}[h!]
\centering
\includegraphics[width=8.0cm,clip=true]{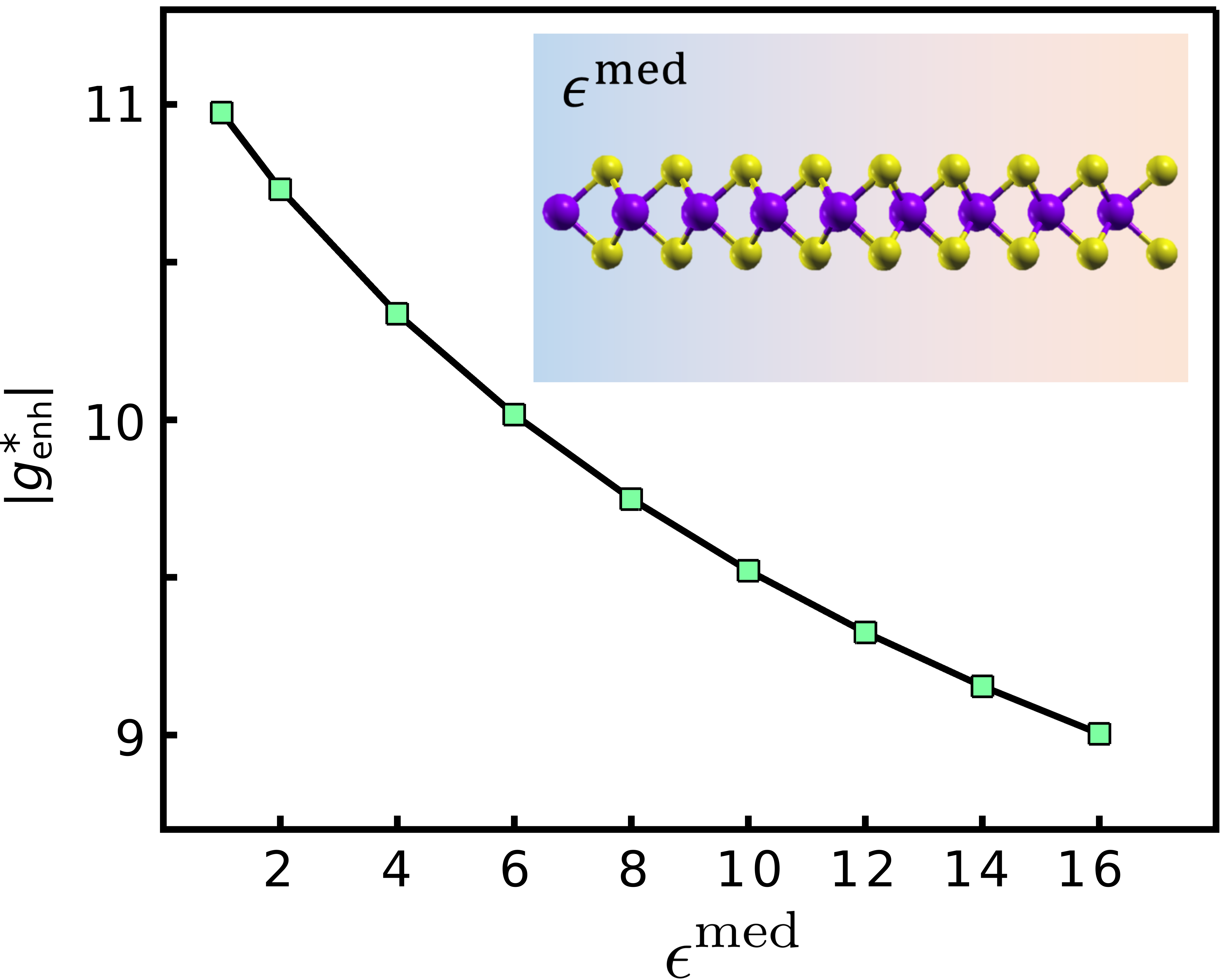}
\caption{
\textbf{Tunability of} $|g^*_{\text{enh}}|$ \textbf{using dielectric screening.}
\textit{Ab initio} valence band g-factor $|g^*_{\text{enh}}|$ as a function of background dielectric constant $\epsilon^{\text{med}}$ for a hole density of $5.18 \times 10^{12}\text{cm}^{-2}$.
}
\label{epsfig}
\end{figure}

The alignment of LLs in different valleys can in principle be achieved for $B<B_c$ if $g^*_{\text{enh}}$ can be tuned such that $0K'$ matches exactly with $NK$ for some $N$. 
However, once $g^*_{\text{enh}}$ deviates slightly from this value, due to fluctuations in the hole density or dielectric environment (see Fig.~\ref{epsfig}), the LLs are no longer aligned. 
Our predictions above enable the alignment of LLs while allowing for some fluctuations in $g^*_{\text{enh}}$. 

\textbf{Tunability of interaction-enhanced g-factor} 

We further note that in addition to electrostatic gating which changes the carrier concentration and thus $g^*_{\text{enh}}$ (Fig.~\ref{gstarfig}), 
$g^*_{\text{enh}}$ can also be tuned by dielectric screening (Fig.~\ref{epsfig}). 
The tunability of $g^*_{\text{enh}}$ with the background dielectric constant can be understood from the fact that $g^*_{\text{enh}}$ is related to the effective quasi-2D screened Coulomb potential at the Fermi surface (Eq.~\ref{enh_main}). 
This tunability of $g^*_{\text{enh}}$ provides a handle to control the valley-polarized current, $B_c$ and $\Delta B$.

\textbf{DISCUSSION}

In summary, our \textit{ab initio} calculations show that many-body interactions in doped TMD MLs enhance the g-factors compared to the undoped MLs, 
up to a critical magnetic field $B_c$ above which the g-factors revert to those in the undoped systems. 
Such a phenomenon has not been predicted or observed in silicon and other traditional valleytronic materials, 
because the corresponding $B_c$ would be much larger due to the smaller g-factors in these materials. 

The enhancement in g-factors arises from the effect of a magnetic-field-induced change in occupancies on the screened exchange interactions. 
This effect is only present in the mixed-polarized regime ($B<B_c$). As the carrier concentration increases, $g^*$ decreases and $B_c$ increases, 
so that for any value of the magnetic field $B_0$, the valley-polarization is maximized when the carrier concentration is such that $B_c = B_0$. 
This prediction has implications for maximizing the valley- and spin-polarized current for the valley Hall effect. 

The computed interaction-enhanced g-factors agree well with experiment for different doping concentrations. We further identify the values of $g^*_{\text{enh}}$ and corresponding ranges of $B$ that lead to a valley-filling instability and expected LL alignment, 
which are of interest for the investigation of quantum phase transitions in doped TMDs \cite{Braz,Donk,Miserev,Roch,Nature1999,Science2000}. 
Recent observations of fractional quantum Hall states associated with non-abelian anyons in ML WSe$_2$ \cite{FQHWSe2} suggest that 
the creation of pseudo-spinors from a linear combination of valley-aligned LLs can be useful for topological quantum computing applications \cite{QCRMP}.

\textbf{METHODS}

\textbf{Calculation of intrinsic g-factor}

The orbital component of the intrinsic g-factor $g^{\text{orb}}_{n\mathbf{K}}$ is defined as $g^{\text{orb}}_{n\mathbf{K}} \mu_B = m^z_{n\mathbf{K}}$ \cite{Xuan2020}:
\begin{eqnarray}
\mathbf{m}_{n\mathbf{K}}=-\frac{ie}{2\hbar}\ev{\times [H_{\mathbf{k}} - E_{n\mathbf{k}}]}{\partial_{\mathbf{k}} u_{n\mathbf{k}}}|_{\mathbf{k}=\mathbf{K}}.
\label{semim}
\end{eqnarray}
We use the PBE exchange-correlation functional \cite{PBE1996} for the DFT mean-field calculations \cite{QE} and the details follow those in Ref. \onlinecite{Xuan2020}.
For $GW$ calculations of the intrinsic g-factors, we use a non-uniform sampling method \cite{sub} of the Brillouin Zone
starting with a $12 \times 12$ k-grid as implemented in the BerkeleyGW code \cite{BGW}.
The energy-dependence of the dielectric function is treated within the generalized plasmon pole (GPP) model \cite{Louie}.
An energy cutoff of $35$ Ry with $4000$ empty bands is used for the reciprocal space expansion of the dielectric matrix.
The intrinsic single-band g-factor reduces by only $0.2$ when an energy cutoff of $2$Ry with $200$ empty bands is used.

\textbf{Calculation of renormalized g-factor}

The renormalized g-factor $g^{*}$ is computed from the intrinsic g-factor $g^I$ and $\frac{d\Sigma(E)}{dE}$ using Eq.~\ref{effg}. 
As the dependence of $E_{n\mathbf{k}}^{\text{QP}}$ on $B$ is no longer linear, 
we generalize Eq.~(\ref{EQP1}) to the case where $E_{n\mathbf{k}}^{\text{QP}}$ refers to the quasiparticle energies in the presence of a $B$-field, 
and the applied $B$ represents a small increment in $B$. 

We approximate $g^I$ using the value in the undoped system. The intrinsic single band g-factors from DFT calculations do not change when ML WSe$_2$ is doped with holes. 

The first term of Eq.~\ref{dSdE} can be obtained directly from the BerkeleyGW output\cite{BGW}. 

$\Sigma=iGW$ can be partitioned \cite{Louie} into the dynamical non-local screened-exchange (SEX) and Coulomb-hole (COH) interaction terms $\Sigma = \Sigma^{\text{SEX}} + \Sigma^{\text{COH}}$. 
Only the screened exchange term depends on the occupancies $f$ and contributes to the second term of Eq.~\ref{dSdE}.
The screened exchange energy $\Sigma^{\text{SEX}}$ in our \textit{ab initio} plane-wave calculation can be written as (see Supplemental Note):
\begin{eqnarray} 
\begin{split}
\Sigma_{n\mathbf{K}}^{\text{SEX}}(E) 
&= -\sum_{m}\frac{1}{N_{\mathbf{q}}\Omega}\sum_{\mathbf{qGG'}} f_{m\mathbf{K}-\mathbf{q}} \bra{n\mathbf{K}}e^{i(\mathbf{q}+\mathbf{G})\cdot\mathbf{r}}\ket{m\mathbf{K}-\mathbf{q}} 
\bra{m\mathbf{K}-\mathbf{q}}e^{-i(\mathbf{q}+\mathbf{G}')\cdot\mathbf{r}}\ket{n\mathbf{K}}\\
&\times \epsilon^{-1}_{\mathbf{G}\mathbf{G}'}(\mathbf{q},E-E_{m\mathbf{K}-\mathbf{q}})v_{\mathbf{q}+\mathbf{G}} \\
&= -\sum_{m} \frac{1}{(2\pi)^2} \int_{BZ} d^2q \bar{W}_{m\mathbf{q}} f_{m\mathbf{K}-\mathbf{q}},
\label{SX}
\end{split}
\end{eqnarray}
where we define the quasi-2D screened Coulomb potential:
\begin{eqnarray}
\begin{split}
\bar{W}_{m\mathbf{q}}(E)  &=  \frac{1}{L}\sum_{\mathbf{GG'}} \bra{n\mathbf{K}}e^{i(\mathbf{q}+\mathbf{G})\cdot\mathbf{r}}\ket{m\mathbf{K}-\mathbf{q}} 
\bra{m\mathbf{K}-\mathbf{q}}e^{-i(\mathbf{q}+\mathbf{G}')\cdot\mathbf{r}}\ket{n\mathbf{K}}\\
&\times \epsilon^{-1}_{\mathbf{G}\mathbf{G}'}(\mathbf{q},E-E_{m\mathbf{K}-\mathbf{q}}) v_{\mathbf{q}+\mathbf{G}}.
\label{wbar}
\end{split}
\end{eqnarray}
Here, $\Omega$ is the cell volume,
$N_{\mathbf{q}}$ is the number of q-points and $v_{\mathbf{q}}$ is the Coulomb potential with the slab Coulomb truncation scheme applied \cite{trunc}. 
$\bar{W}_{m\mathbf{q}}$ is an effective quasi-2D screened Coulomb potential defined in valley $K$ and $L$ is the height of the supercell for ML WSe$_2$. 
The second term in Eq.~(\ref{dSdE}) is then given by (see Supplemental Note):
\begin{eqnarray} 
\begin{split}
\frac{\partial\Sigma(E)}{\partial f}\frac{\partial f}{\partial E} 
&= -\sum_{m} \frac{1}{(2\pi)^2} \int_{BZ} d^2q \bar{W}_{m\mathbf{q}} \frac{\partial f_{m\mathbf{K}-\mathbf{q}}}{\partial E}\approx \frac{|m^*|}{2\pi} \bar{W}_{nk_F}(E=E_F).
\label{enh}
\end{split}
\end{eqnarray}

We compute $g^*_{v\mathbf{K}}$ by evaluating $\bar{W}_{vk_F}$ \textit{ab initio} using the random phase approximation for the dielectric matrix. 
Due to the partial occupancies, we calculate the dielectric matrix using a dense reciprocal space sampling of $120\times120$, a $2$Ry $G$-vector cut off and $29$ bands.
$g^*$ is unchanged when we use instead $4$Ry and $299$ bands, and reduces by $\sim 3 \%$ when a $240\times240$ k-mesh is used.
Care is taken to include the effect of spin-orbit splitting at the valleys.
For the effective mass, we use our DFT value of $m^*=-0.48m_e$, which agrees well with the experimentally deduced value for hole-doped WSe$_2$ \cite{LL2016}. 
If electronic screening is ignored, the effective quasi-2D bare Coulomb potential $\bar{V}_{mq}$ is defined by:
\begin{eqnarray}
\bar{V}_{mq}(E)  =  \frac{1}{L}\sum_{\mathbf{GG'}} \bra{n\mathbf{k}}e^{i(\mathbf{q}+\mathbf{G})\cdot\mathbf{r}}\ket{m\mathbf{k}-\mathbf{q}} 
\bra{m\mathbf{k}-\mathbf{q}}e^{-i(\mathbf{q}+\mathbf{G}')\cdot\mathbf{r}}\ket{n\mathbf{k}}
\delta_{\mathbf{G}\mathbf{G}'} v_{\mathbf{q}+\mathbf{G}}.
\label{wbar}
\end{eqnarray}
Our first principles results for $\bar{V}_{vk_F}$ (Supplementary Figure 1) agrees with the analytically-derived 2D Coulomb potential.
At low doping densities, $\bar{V}_{vk_F}$ is very large, which would change the sign of $g^*$ compared to the intrinsic g-factor,
indicating that screening is important for a meaningful description of $g^*$.

\textbf{Background dielectric constant}

A uniform background dielectric constant ($\epsilon^{\text{med}}$) can be simply added to the  dielectric function of the system to obtain the total dielectric function: 
$\epsilon(\mathbf{r},\mathbf{r}',\omega) = \epsilon^{\text{WSe}_2}(\mathbf{r},\mathbf{r}',\omega) + \epsilon^{\text{med}}-1$.
In our first principles calculation, the dielectric function is expanded in a plane wave basis:
$\epsilon(\mathbf{r},\mathbf{r}',\omega) = \sum_{\mathbf{qGG'}} e^{i(\mathbf{q}+\mathbf{G})\cdot\mathbf{r}}\epsilon_{\mathbf{G}\mathbf{G}'}(\mathbf{q},\omega) e^{-i(\mathbf{q}+\mathbf{G}')\cdot\mathbf{r}'}$.
Thus we approximate the effect of screening by a dielectric medium by modifying the static dielectric matrix as follows:
\begin{eqnarray}
\epsilon_{\mathbf{G}\mathbf{G}'}(\mathbf{q},0) = \epsilon^{\text{WSe}_2}_{\mathbf{G}\mathbf{G}'} + (\epsilon^{\text{med}}-1) \delta_{\mathbf{G}\mathbf{G}'}.
\end{eqnarray}

\textbf{ACKNOWLEDGEMENTS}

This work is supported by the NUS Provost's Office, the Ministry of Education (MOE 2017-T2-2-139) and the National Research Foundation (NRF), 
Singapore, under the NRF medium-sized centre programme. 
Calculations were performed on the computational cluster in the Centre for Advanced 2D Materials and the National Supercomputing Centre, Singapore.

\clearpage

\begin{center}
\bf{Supplementary Information for:} 
\bf{Valley-Filling Instability and Critical Magnetic Field for Interaction-Enhanced Zeeman Response in Doped WSe$_2$ Monolayers}
\end{center}

\clearpage

%\section{Supplementary Figures}
%\begin{figure}[!h]
%\centering
%\includegraphics[width=8.9cm,clip=true]{SIVbar.png}
%\caption{
%\textbf{Quasi-2D bare Coulomb potential $\bar{V}_{vk_F}$ as a function of hole density.}
%Black squares: first principles results in atomic units computed using the valence band wavefunctions at $K$.
%Blue curve: analytical 2D Coulomb potential.
%[XFY: need to truncate to 7? Use the same x-axis range as figure for Wbar.]
%}
%\end{figure}
%
\begin{figure}[!h]
\centering
\includegraphics[width=17.0cm,clip=true]{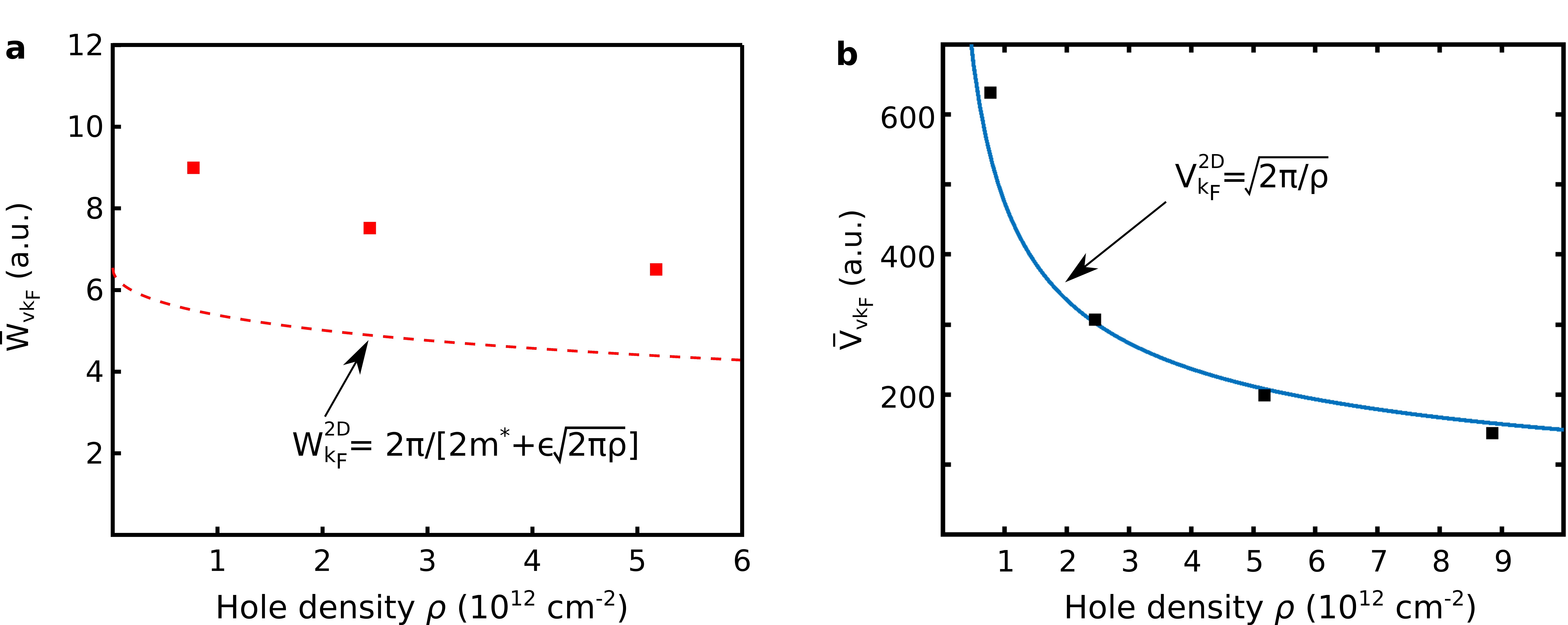}
\caption{
\textbf{Quasi-2D screened Coulomb potential $\bar{W}_{vk_F}$ and bare Coulomb potential $\bar{V}_{vk_F}$ as a function of hole density.} 
\textbf{a} Red squares: first principles results for $\bar{W}_{vk_F}$ in atomic units computed using the valence band wavefunctions at $K$.
Dashed line: screened Coulomb potential for an ideal 2D fermion gas \cite{stern} using a dielectric constant $\epsilon$ of $15.6$ \cite{medeps} for WSe$_2$.
\textbf{b} Black squares: first principles results for $\bar{V}_{vk_F}$ in atomic units computed using the valence band wavefunctions at $K$.
Blue curve: analytical 2D bare Coulomb potential.
}
\end{figure}

\begin{figure}[!h]
\centering
\includegraphics[width=9.1cm,clip=true]{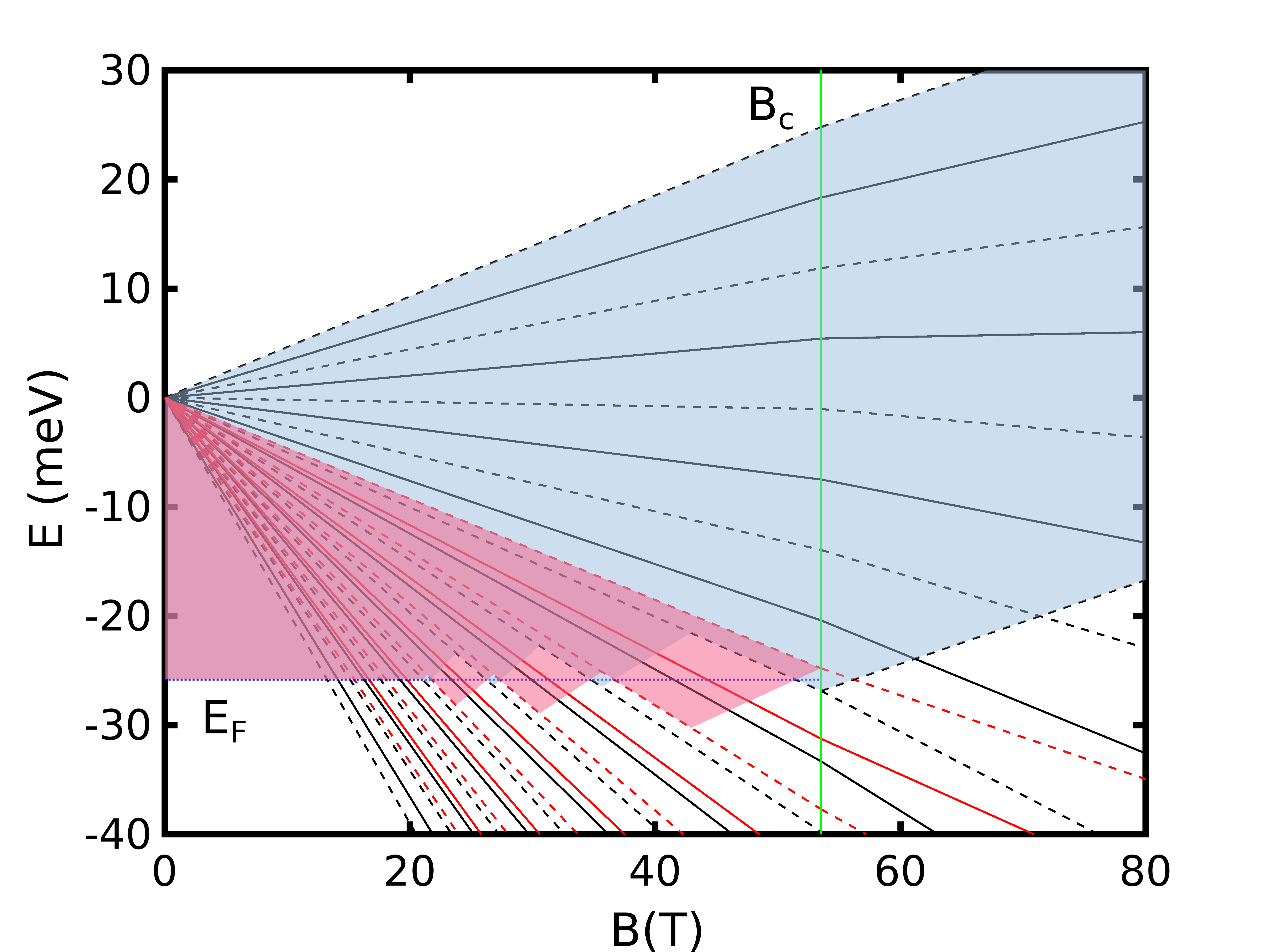}
\caption{
\textbf{LL fan diagram for ML WSe$_2$ valence band with a hole density of} $5.2\times 10^{12}\text{cm}^{-2}$\textbf{.} Here, $g^*_{\text{enh}} = -8.0$ and no valley-filling instability occurs.
}
\end{figure}

\begin{table}
\caption{
\textbf{Values of} $B_c$, $B_X$ \textbf{and} $\Delta B$ (T) \textbf{for given} $|g^*_{\text{enh}}|$\textbf{.}
}
\begin{ruledtabular}
\begin{tabular}{l*{4}{c}}
$|g^*_{\text{enh}}|$           & $B_c$ & $B_X$ & $B_X+\Delta B$ & $\Delta B$\\
\hline
$[10.4, 11.2]$        &  $[38.5, 31.2]$  & $[38.5, 37.4]$  & $[46.2, 37.4]$  & $[7.7, 0]$\\
$[12.5, 13.7]$        &  $[16.7, 10.0]$  & $[16.7, 11.7]$  & $[19.5, 11.7]$  & $[2.8, 0]$ \\
\end{tabular}
\label{tabB}
\end{ruledtabular}
\end{table}

\section{Supplementary Note}

\subsection{Derivation of the expression for $\frac{\partial\Sigma(E)}{\partial f}\frac{\partial f}{\partial E}$}

The screened exchange term in our \textit{ab initio} $GW$ self-energy can be written as:
\begin{eqnarray} 
\begin{split}
\Sigma^{\text{SEX}}_{n\mathbf{K}}(E) &= -\sum_{m}\frac{1}{N_\mathbf{q} \Omega}\sum_{\mathbf{qGG'}} f_{m\mathbf{K}-\mathbf{q}} 
\bra{n\mathbf{K}}e^{i(\mathbf{q}+\mathbf{G})\cdot\mathbf{r}}\ket{m\mathbf{K}-\mathbf{q}} 
\bra{m\mathbf{K}-\mathbf{q}}e^{-i(\mathbf{q}+\mathbf{G'})\cdot\mathbf{r'}}\ket{n\mathbf{K}} \\
&\times\epsilon^{-1}_{\mathbf{G}\mathbf{G}'}(\mathbf{q},E-E_{m\mathbf{K}-\mathbf{q}})v_{\mathbf{q}+\mathbf{G}} \\
&= -\sum_{m}\frac{1}{L}\frac{1}{N_\mathbf{q} S}\sum_{\mathbf{qGG'}} f_{m\mathbf{K}-\mathbf{q}} \bra{n\mathbf{K}}e^{i(\mathbf{q}+\mathbf{G})\cdot\mathbf{r}}\ket{m\mathbf{K}-\mathbf{q}} 
\bra{m\mathbf{K}-\mathbf{q}}e^{-i(\mathbf{q}+\mathbf{G'})\cdot\mathbf{r'}}\ket{n\mathbf{K}} \\
&\times\epsilon^{-1}_{\mathbf{G}\mathbf{G}'}(\mathbf{q},E-E_{m\mathbf{K}-\mathbf{q}})v_{\mathbf{q}+\mathbf{G}} \\
&= -\sum_{m} \frac{1}{(2\pi)^2} \int d^2q \frac{1}{L}\sum_{\mathbf{GG'}}f_{m\mathbf{K}-\mathbf{q}} \bra{n\mathbf{K}}e^{i(\mathbf{q}+\mathbf{G})\cdot\mathbf{r}}\ket{m\mathbf{K}-\mathbf{q}} 
\bra{m\mathbf{K}-\mathbf{q}}e^{-i(\mathbf{q}+\mathbf{G'})\cdot\mathbf{r'}}\ket{n\mathbf{K}}\\
&\times\epsilon^{-1}_{\mathbf{G}\mathbf{G}'}(\mathbf{q},E-E_{m\mathbf{K}-\mathbf{q}})v_{\mathbf{q}+\mathbf{G}} \\
&= -\sum_{m} \frac{1}{(2\pi)^2} \int_{BZ} d^2q \bar{W}_{m\mathbf{q}} f_{m\mathbf{K}-\mathbf{q}},
\end{split}
\end{eqnarray}
where $L$ is the cell height, $S$ is the cell area, $\Omega$ is the cell volume, $N_\mathbf{q}$ is the number of q points.

In the above expression, the inverse dielectric function is written in a reciprocal space basis 
\begin{eqnarray}
\epsilon^{-1}(\mathbf{r},\mathbf{r}',E)=\frac{1}{\Omega}\sum_{\mathbf{qGG'}}e^{i(\mathbf{q}+\mathbf{G})\cdot\mathbf{r}}
\epsilon^{-1}_{\mathbf{G}\mathbf{G}'}(\mathbf{q},E-E_{m\mathbf{K}-\mathbf{q}})
e^{-i(\mathbf{q}+\mathbf{G'})\cdot\mathbf{r'}},
\end{eqnarray}
and $\bar{W}_{m\mathbf{q}}(E)$ is defined as
\begin{eqnarray}
\begin{split}
\bar{W}_{m\mathbf{q}}(E)  &=  \frac{1}{L}\sum_{\mathbf{GG'}} \bra{n\mathbf{K}}e^{i(\mathbf{q}+\mathbf{G})\cdot\mathbf{r}}\ket{m\mathbf{K}-\mathbf{q}} 
\bra{m\mathbf{K}-\mathbf{q}}e^{-i(\mathbf{q}+\mathbf{G'})\cdot\mathbf{r'}}\ket{n\mathbf{K}}\\
&\times\epsilon^{-1}_{\mathbf{G}\mathbf{G}'}(\mathbf{q},E-E_{m\mathbf{K}-\mathbf{q}}) v_{\mathbf{q}+\mathbf{G}}.
\label{wbar}
\end{split}
\end{eqnarray}

$\frac{\partial\Sigma(E)}{\partial f}\frac{\partial f}{\partial E}$ can be simplified as:
\begin{eqnarray} 
\begin{split}
\frac{\partial\Sigma(E)}{\partial f}\frac{\partial f}{\partial E} &= -\sum_{m} \frac{1}{(2\pi)^2} 
\int d^2q \bar{W}_{m\mathbf{q}} \frac{\partial f_{m\mathbf{K}-\mathbf{q}}}{\partial E}\\
&= \sum_{m} \frac{1}{(2\pi)^2} \int d^2q \bar{W}_{m\mathbf{q}}\delta(E_{m\mathbf{K}-\mathbf{q}} - E_F)\\
&= \frac{1}{(2\pi)^2} \int d^2q \bar{W}_{n\mathbf{q}} \delta(E_{n\mathbf{q}} - E_F) \\
&= \frac{1}{(2\pi)^2} \int qdqd\phi_{q} \bar{W}_{n\mathbf{q}} \delta(E_{\mathbf{q}} - E_F)\\
&= \frac{1}{(2\pi)^2} \int_0^{2\pi} qdqd\phi_{q} \frac{\bar{W}_{n\mathbf{q}}}{|dE_{\mathbf{q}}/dq|}|_{q=k_F}\\ 
&\approx \frac{|m^*|}{2\pi} \bar{W}_{nk_F}(E=E_F),
\end{split}
\end{eqnarray}
where the delta function picks up the integrand at the Fermi surface and $n$ is the band index of the frontier doped band. We set $E = E_F$, which is equivalent to taking the approximation that $\epsilon^{-1}_{\mathbf{G}\mathbf{G}'}(\mathbf{k_F},E_K-E_F)\approx \epsilon^{-1}_{\mathbf{G}\mathbf{G}'}(\mathbf{k_F},0)$, where $E_K$ is the energy at the band extremum in $K$. This approximation is valid because $E_K-E_F$ is on the order of meVs, and the dielectric function is fairly constant in this energy range. \cite{LiYang}

\subsection{Critical magnetic field $B_c$ and condition for valley-filling instability}
The condition for the valley-filling instability is given by $\Delta B > 0$. In the following, we derive expressions for $B_c$, $B_X$, and $\Delta B$. 
Let $NK$ be the $K$-valley LL so that the energy of $0K'$ lies in between those of $NK$ and $(N+1)K$ when $B<B_c$. This is equivalent to the condition that 
\begin{eqnarray}
\frac{N}{m^*}<|g^*_{\text{enh}}|<\frac{N+1}{m^*}
\label{g*int}
\end{eqnarray}

$B_c$ is defined as the minimum value of $B$ at which $0K'$ is just completely empty with holes. 
This implies that $NK$ must also be completely occupied with holes at $B=B_c$. All the hole population originally in $0K'$ for $B<B_c$ has been transferred to $NK$ at $B=B_c$. Thus, at $B=B_c$, we have: 
\begin{eqnarray}
\epsilon_{0K'}+\frac{\hbar \omega_c}{2}-E_F = E_F-(\epsilon_{NK}-\frac{\hbar \omega_c}{2})
\end{eqnarray}
So, 
\begin{eqnarray}
\frac{\epsilon_{NK}+\epsilon_{0K'}}{2}=E_F
\end{eqnarray}
where $E_F$ is the position of the Fermi level in the absence of the magnetic field. This condition leads to the expression:
\begin{eqnarray}
B_c = \frac{|m^*||E_F|}{(N+1)\mu_B}.
\end{eqnarray}
This expression for $B_c$ gives results very similar to $B_c \sim |E_F|/(|g^*_{\text{enh}}|\mu_B)$ given in the main text.

$B_X$ is obtained by finding the value of $B$ at which  $\epsilon_{NK}=\epsilon_{0K'}$, for $B>B_c$:  
\begin{eqnarray}
B_X = \frac{(|g^*_{\text{enh}}|-|g^*_0|)B_c}{N/|m^*|-|g^*_0|},
\end{eqnarray}
where $g^*_0$ is the renormalized g-factor for $B>B_c$, i.e. the renormalized g-factor of the undoped system, and $g^*_{\text{enh}}$ refers to the interaction-enhanced g-factor.
$B_X +\Delta B$ is the value of $B$ at which the $K$-valley LLs from index $0$ to $(N-1)$ can just contain all the holes, 
and can be obtained as the value of $B$ for the crossing point indicated by the yellow circle in Figure 4d of the main text. Thus, we have:
\begin{eqnarray}
B_X +\Delta B= (1+\frac{1}{N})B_c.
\end{eqnarray}
This gives:
\begin{eqnarray}
\Delta B = [\frac{N/|m^*|-|g^*_{\text{enh}}|}{N/|m^*|-|g^*_0|}+\frac{1}{N}]B_c.
\end{eqnarray}
The instability condition for $g^*_{\text{enh}}$ can be obtained from $\Delta B > 0$. Together with Eq.~\ref{g*int}, we have
\begin{eqnarray}
\frac{N}{m^*}<|g^*_{\text{enh}}|<\frac{N+1}{m^*}-\frac{|g^*_0|}{N}.
\end{eqnarray}
This expression also implies a condition on $N$ given by
\begin{eqnarray}
\frac{N}{m^*}<\frac{N+1}{m^*}-\frac{|g^*_0|}{N}.
\end{eqnarray}
Using our prediction of $|g^*_0|=6.6$ and $m^*=-0.48$, the minimum $N$ is $4$, giving the minimum $|g^*_{\text{enh}}|$ for a valley-filling instability to be $8.33$.

\end{document}